\documentclass{elsarticle}

\usepackage{epic}
\usepackage{psfrag}
\usepackage{latexsym}
\usepackage{amsmath}
\usepackage{epsfig}
\usepackage{amssymb}
\usepackage{enumerate}

\usepackage[mathscr]{euscript}

\newtheorem{theorem}{Theorem}[section]
\newtheorem{lemma}[theorem]{Lemma} 
\newtheorem{corollary}[theorem]{Corollary}
\newtheorem{proposition}[theorem]{Proposition}

\newcommand{\PP}{{\mathbb P}}
\newcommand{\EE}{{\mathbb E}}

\newcommand{\cI}{{\mathcal I}}
\newcommand{\cE}{{\mathcal E}}
\newcommand{\cF}{{\mathcal F}}
\newcommand{\cT}{{\mathcal T}}

\journal{Theoretical Population Biology}
\begin{document}
\begin{frontmatter}

\title{Clades, clans and reciprocal monophyly under neutral evolutionary models}
\author{Sha Zhu, James H. Degnan, Mike Steel}
\address{Biomathematics Research Centre, University of Canterbury, Private Bag 4800, Christchurch 8140, New Zealand}

\date{\today}
\begin{abstract}
The Yule model and the coalescent model are two neutral stochastic models for generating trees in phylogenetics and population genetics, respectively. Although these models are quite different, they lead to identical distributions concerning the probability that
pre-specified groups of taxa form monophyletic groups (clades) in the tree. We extend earlier work to derive exact formulae for the probability of
finding one or more groups of taxa as clades in a rooted tree, or as `clans' in an unrooted tree.  Our findings are relevant for calculating the statistical significance of observed monophyly and reciprocal monophyly in phylogenetics.
\end{abstract}


\end{frontmatter}

\newpage
\section{Introduction}
\label{Introduction}

When gene trees are estimated from multiple lineages taken from two or more populations, there is an increased chance that lineages within each population form monophyletic groups compared to sampling multiple lineages from a single population.  This observation has led to the adoption of a null hypothesis that a set of lineages belongs to a single population or taxonomic group, in asking whether a particular group of lineages came from a taxonomically distinct population \cite{rosen07, cum}. Statistical tests for reciprocal monophyly between two sister taxa can then be developed to test against this null hypothesis  \cite{hudson2002, rosen03}.

Reciprocal monophyly is central to the genealogical species concept. According to this concept two groups come from different species if they form distinct monophyletic groups \cite{deQuieroz2007, hudson2002}.  Gene trees from lineages sampled from one or more populations are typically estimated, and monophyly (or lack of monophyly) of these groups can be observed from the clades of the gene tree.  Statistical tests for whether observed levels of monophyly provide sufficient evidence to conclude that a group is taxonomically distinct can be performed, given a probabilistic model for the clades on a tree \cite{rosen07}.

Two neutral models -- involving different evolutionary scales -- are useful in this context.   The Yule (pure birth, or birth-death) model describes the speciation (and extinction) of lineages at the species level as one moves forward in time,
 while Kingman's coalescent process is a population genetic model that the ancestry of individual lineages back in time as they coalesce (and thereby form a tree).   These are two quite different processes and lead to different branch lengths on trees; remarkably, however, they generate identical distributions of tree topologies \cite{ald}.  Thus, while the coalescent process is a natural model for trees in single populations, the equivalence of the Yule and coalescent models for tree topologies means that results for the Yule model can be exploited in studying probabilities of clades for coalescent trees in single populations.

Although there has been an emphasis on testing for the taxonomic distinctiveness of one group of lineages, joint probabilities of clades could be used to examine whether the observed monophyly of several groups is statistically significant using a single test.  Such an omnibus test of the null hypothesis that all groups come from one population might be more powerful than testing several groups one at time.

In this note, we derive exact formulae for the joint probabilities of $k$ clades for a random Yule/coalescent gene tree under the conditions that the $k$ clades are mutually exclusive (they have no leaves of the gene tree in common), and are either exhaustive (all leaves of the gene tree occur in one of the $k$ clades), or form only a subset of the leaves of the gene tree.   These results generalize results from \cite{rosen03}, which provided an explicit formula for the probability that two mutually exclusive and exhaustive sets of leaves formed clades on a Yule/coalescent gene tree.

In addition, we extend the results to unrooted trees by giving the probabilities of `clans' (sets of leaves that are all on one side of a split \cite{wilk}), as well as the joint probability of $k>1$ clans, on Yule/coalescent trees which have been unrooted.   This extension is relevant when only unrooted trees can be estimated, which is particularly common in prokaryotic evolution \cite{lap}.

\section{Clades}

\begin{figure}[ht]
\psfrag{PSFRAG_LABEL1}{$\rho$}
\begin{center}
\resizebox{12cm}{!}{
\includegraphics{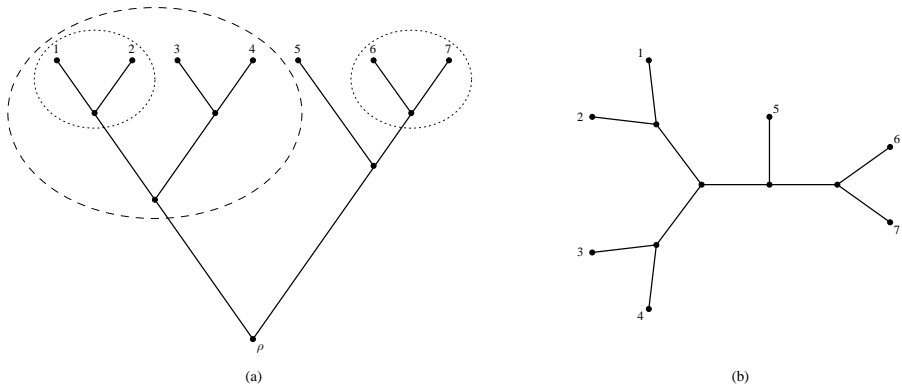}
}
\end{center}
\caption{(a) This rooted tree has 13 `clades', including the three sets circled ($\{1,2\}, \{1,2,3,4\}, \{6,7\})$. In this tree $\{1,2\}$ and $\{3,4\}$ are sister clades, but $\{1,2\}$ and $\{6, 7\}$ are not.  (b) The unrooted tree $T^{-\rho}$ obtained from the tree $T$ in (a) by suppressing the root vertex $\rho$. This tree has
$\{3,4,5,6,7\}$ as a `clan', even though this set is not a clade of $T$.}
\label{fig1}
\end{figure}

Throughout this paper we will let $X_n$ (or, more briefly, $X$) denote a set of taxa of size $n$.
Given a rooted phylogenetic $X$--tree $T_X$ (more briefly $T$), with leaf set $X=X_n$,  a {\em clade  of $T$} is a subset of $X$ that corresponds to the set of leaves that are descended from any internal vertex. For example, in
Fig.~\ref{fig1}(a), the sets $\{3,4\}$ and $\{1,2,3,4\}$ are two clades. Any two clades $A$ and $B$ of $T$ satisfy the following {\em compatibility} condition:
\begin{equation}
\label{nesteq}
A \cap B \in \{A, B, \emptyset\}.
\end{equation}
This is equivalent to requiring that $A=B$, one set is a strict subset of the other, or the two sets are disjoint.

We will let $c(T)$ denote the set of clades of $T$, and say that
a clade is {\em proper} if it is a strict subset of $X$.  Notice that a rooted phylogenetic tree $X$--tree has at most $2n-1$ clades, and it has precisely this number if and only the tree
is {\em binary}, that is, if each non-leaf vertex has two descendant vertices.

\section{The Yule-Harding-Kingman process}
\begin{figure}[ht]
\resizebox{12cm}{!}{
\includegraphics{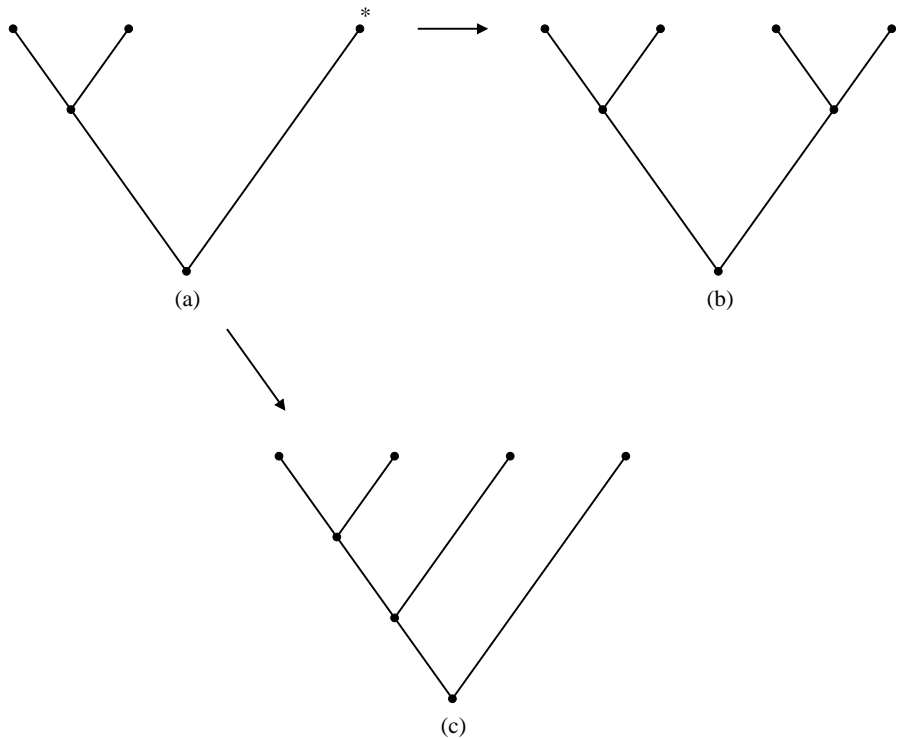}
}
\caption{From a rooted binary tree on three leaves (a), splitting the right leaf (*) leads to a `balanced' tree shape (b), while splitting either of the other two leaves produces an unbalanced tree (c). Thus the balanced tree shape has probability $1/3$ and as there are three distinct ways to label the leaves, each of these rooted binary phylogenetic trees has probability $1/9$ under the YHK process. For a phylogenetic tree of shape (c), the probability is $1/18$.}
\label{fig2}
\end{figure}

Consider the probability distribution on binary phylogenetic $X$--trees described by a model that grows a tree by selecting a leaf uniformly at random and `splitting' it into two new leaves, as illustrated in Fig.~\ref{fig2}.  Since we are ignoring branch lengths in this paper and concentrating just on tree topologies,  the resulting probability distribution on rooted binary tree topologies is the same as  that given by any (stationary or non-stationary) birth-death process on trees in which birth (speciation) and death (extinction) events apply exchangeably to all the species extant at any
given moment (see \cite{ald} for further details).  This is useful, since the rates of speciation and extinction throughout time may be both time-dependent and variable according to the number of taxa present \cite{rab}.

The study of such pure-birth trees was initiated in Yule's 1925 paper \cite{yule}, and the probability distribution on tree topologies (without reference to branch lengths) was further studied by Harding  \cite{har}.
Moreover, this probability distribution on trees is  precisely the same as that given by a quite different process, namely
Kingman's coalescent process \cite{king} in population genetics, which starts at the leaves and successively combines pairs of elements, provided that, once again, we ignore branch lengths (\cite{ald}).

To emphasize this equivalence between a model in macro-evolution (speciation and extinction) and micro-evolution (population genetics) we will refer to it as the {\em Yule-Harding-Kingman (YHK)} process for generating tree topologies.

  We will also refer to a random binary phylogenetic $X$--tree  produced by any of these stochastically equivalent processes as $\cT_X$ (or often just $\cT$ if $X$ is clear), and so
$\PP(\cT_X = T)$ is the probability that $T$ is the actual phylogenetic $X$--tree produced by the process.  The process, viewed as a pure-birth model, is illustrated in Fig.~\ref{fig2}.

In this paper, we exploit two important properties of the process that generates $\cT$.
First we recall some notation that will be used throughout: for any phylogenetic $X$--tree and any non-empty subset $Y$ of $X$, let $T_{X|Y}$ be the phylogenetic tree induced by restricting the leaf set to $Y$ (as in \cite{sem}).
The two properties that the YHK process enjoys, and which we will exploit throughout this paper, are the following:

\begin{itemize}
\item[{\bf (EP)}] If $T'$ is obtained from $T$ by permuting its leaves, then $$\PP(\cT = T') = \PP(\cT=T).$$
\item[{\bf (GE)}]  For any proper (and non-empty) subset $A$ of $X$, and any  rooted binary phylogenetic tree $T$ with leaf set $X-A$:  $$\PP(\cT_{X|(X-A)} = T | A \in c(\cT)) = \PP(\cT_{(X-A)} = T).$$
\end{itemize}
Property (EP) is the  {\em Exchangeability} property \cite{ald}, which requires that the probability of a particular phylogenetic tree depends just on its  shape and not on how its leaves are labeled (it is called
`label-invariance' in \cite{penny}).
Property  (GE) is  the {\em Group Elimination} property from \cite{ald};  it states that, conditional on $A$ forming a clade in the tree, the
tree structure on the remaining taxa is also described by the YHK process. In turn (GE) implies the following  {\em Sampling Consistency}  property (\cite{ald}):
 For any rooted binary tree $T$ with leaf set $A \subseteq X$, we have:
 \begin{itemize}
\item[{\bf (SC)}] $\PP(\cT_{X|A} = T) = \PP(\cT_A = T).$
\end{itemize}
To see that (GE) implies (SC), one sequentially deletes leaves that are not in $A$, noting that each leaf is, trivially, a clade in any tree.

\section{Clade probabilities under the YHK process}

The following result is stated and established in the appendix of \cite{heard}; it is also stated and proved in \cite{rosen06} (Theorem 4.4),  and in  \cite{blum} (Proposition 2).
A further proof of this result is also possible based on induction on $n$ and using the well-known property of the YHK model that the number of leaves in one of the (randomly selected)
 maximal subtrees of $\cT_X$  is uniformly distributed between $1$ and $n-1$.

\begin{lemma}
\label{early}
Let $X_n(a)$ be the number of proper clades of size $a$ in $\cT_X$.   Then $$\EE[X_n(a)] = \frac{2n}{a(a+1)}, \mbox{   } 1 \leq a \leq n-1.$$
\end{lemma}

For a subset $A$ of $X$, let $p_n(A)$ be the probability that $A$ is a proper clade of $\cT_X$.   From (EP) it is clear that this probability depends only on $a=|A|$ and $n$, and so we can write $p_n(a)$ for this probability.
From \cite{rosen03} we have:
\begin{lemma}
\label{lemclus}
$$p_n(a) = \begin{cases}
 \frac{2n}{a(a+1)}\binom{n}{a}^{-1}, &\text{ if } 1 \leq a \leq n-1;\\
 0, &\text{ otherwise. }
\end{cases}
 $$
\end{lemma}
The proof of this result from \cite{rosen03} relies on a combinatorial identity to sum a series. Here we point out how Lemma~\ref{lemclus} follows very directly  from Lemma~\ref{early}.

{\em Proof of Lemma~\ref{lemclus}:}
For $1 \leq a \leq n-1$, the exchangeability property (EP) implies that:
$$p_n(A) = \sum_{k \geq 0}\PP(\cT \mbox{ has $k$ clades of size } a) \cdot \frac{k}{\binom{n}{a}} = \EE[X_n(a)]\binom{n}{a}^{-1},$$
where $X_n(a)$ is as defined in Lemma~\ref{early}.  This completes the proof.
\hfill$\Box$

\subsection{Pairs of clades}

For a pair $A,B$ of disjoint subsets  of $X$, let $\hat{p}_n(A,B)$ be the probability that $A$ and $B$ are {\em sister clades} of $\cT_X$ (i.e. $A,B$ and $A \cup B$ are clades of $\cT_X$).  By exchangeability (EP),
this probability depends on $a=|A|, b=|B|$ and $n$ only, and so we will denote it $\hat{p}_n(a,b)$.

Consider first the special case where $n=a+b$; that is,  $A$ and $X-A$ are sister clades, which is equivalent to saying that $A$ is a maximal proper clade.  From \cite{brown} (Equation 6) (see also \cite{rosen03}), the probability of this event is given as follows:
\begin{lemma}
\label{baslem}
For $1 \leq a \leq n$, we have:
$$\hat{p}_n(a,n-a) = \frac{2}{n-1}\binom{n}{a}^{-1}.$$
\end{lemma}

We generalize this slightly as follows:
\begin{lemma}
\label{helpslem}
Let $k=a+b \leq n$. Then:
$$\hat{p}_n(a,b) = \frac{4a!b!(n-k)!}{(n-1)!k(k^2-1)}.$$
\end{lemma}
{\em Proof:}
$$\hat{p}_n(A,B) = \PP(A\cup B \in c(\cT_X)) \cdot \PP\left(A \in c(\cT_{X|A\cup B})|A \cup B \in c(\cT_X)\right).$$
Applying Lemma~\ref{lemclus} to the first term, and property (SC) and Lemma~\ref{baslem} to the second term we have:
$$\hat{p}_n(A,B) = \frac{2n}{(a+b)(a+b+1)}\binom{n}{a+b}^{-1}\cdot \frac{2}{a+b-1}\binom{a+b}{a}^{-1},$$
from which the result follows.
\hfill$\Box$

\bigskip

Now, for any two arbitrary subsets $A,B$ of $X_n =\{1, \ldots, n\}$, let $p_n(A,B)$ be the probability that a Yule tree $\cT$ on $X_n$ has
$A$ and $B$ as proper clades.    As usual, let $a=|A|$ and $b=|B|$.

\begin{theorem}
\label{main}
$$p_n(A,B)= \begin{cases}
p_n(a) &\text{ if } A=B \mbox{ \rm{ [case 1]} };\\
R_n(a,b),   &\text{ if } A \subsetneq B \mbox{ \rm{ [case 2] }};\\
R_n(b, a), &\text{ if } B \subsetneq A \mbox{\rm{ [case 3] }};\\
\hat{p}_n(a, n-a),  &\text{ if } A \cap B = \emptyset, A \cup B = X_n \mbox{\rm{ [case 4]} };\\
r_n(a,b),  &\text{ if } A \cap B = \emptyset, A \cup B \subsetneq X_n \mbox{ \rm{[case 5] }};\\
0,  &\text{ otherwise \mbox{\rm{ [case 6] }}; }
\end{cases}
$$
where $$p_n(a), \mbox{ and }  \hat{p}_n(a,n-a) \mbox{ are given by Lemmas~\ref{lemclus} and \ref{baslem}},$$
$$R_n(a,b):= \frac{4n}{a(a+1)(b+1)}\binom{n}{b}^{-1}\binom{b}{a}^{-1},$$
 $$r_n(a,b) := \frac{4a!b!(n-a-b)!}{(n-1)!}G_n(a,b), \mbox{ {\rm and where}} $$
$$G_n(a,b) := \frac{n}{ab(a+1)(b+1)} - \frac{a(a+1)+b(b+1)+ab}{ab(a+1)(b+1)(a+b+1)} + \frac{1}{(a+b)((a+b)^2-1)}.$$
\end{theorem}
{\em Proof:}
Cases 1 and 4 are given by Lemmas~\ref{lemclus} and \ref{baslem}, respectively.  For the  second case ($A \subsetneq B$), we have:
$$p_n(A,B) = \PP(A \in c(\cT_X)|B \in c(\cT_X)) \cdot \PP(B \in c(\cT_X)).$$
Since $A \subsetneq B$ we can apply property (SC) and Lemma~\ref{lemclus} to deduce that the first term in this product is
$\frac{2b}{a(a+1)}\binom{b}{a}^{-1},$ while the second term is
$\frac{2n}{b(b+1)}\binom{n}{b}^{-1},$ from which the result follows.
Case 3 follows by an analogous argument.
 For Case 5, consider the following two pairs of events:
 \begin{itemize}
 \item $\cE_1: A, B \in c(\cT_X),$
 \item $\cE_2: A\cup B, B \in c(\cT_X),$
 \item $\cF_1:  A \in c(\cT_{X|(X-B)}),$
 \item $\cF_2: B \in c(\cT_X)$.
 \end{itemize}
We are interested in computing $\PP(\cE_1)$  since this is $p_n(A,B)$ and by the principle of inclusion and exclusion we have:
\begin{equation}
\label{pieeq}
\PP(\cE_1) = \PP(\cE_1 \cup \cE_2) + \PP(\cE_1 \cap \cE_2) - \PP(\cE_2).
\end{equation}
Now, $\cE_1 \cup \cE_2$ occurs precisely if $\cF_1 \cap \cF_2$ occurs (since $\cE_1 \cup \cE_2$ is the event that
$B \in c(\cT_X)$ and either $A \in c(\cT_X)$ or $A \cup B \in c(\cT_X)$).  Thus:
$$\PP(\cE_1 \cup \cE_2)  = \PP(\cF_1|\cF_2)\cdot \PP(\cF_2).$$
Combining this equation with (\ref{pieeq}) and noting that $\PP(\cE_1 \cap \cE_2) =  \hat{p}_n(A,B)$ and $p_n(A,B) = \PP(\cE_1)$, we obtain:
\begin{equation}
\label{ppeq2}
p_n(A,B) =\PP(\cE_1)=  \PP(\cF_1|\cF_2)\cdot \PP(\cF_2) - \PP(\cE_2)  + \hat{p}_n(A,B).
\end{equation}
Now, by (GE),
\begin{equation}
\label{ppeq3}
\PP(\cF_1|\cF_2) = \PP(A \in c(\cT_{X-B})) = p_{n-b}(a),
\end{equation}
 and
\begin{equation}
\label{ppeq4}
\PP(\cE_2) = \PP(A \cup B \in c(\cT_X)) \cdot \PP(B \in c(\cT_X)|A \cup B \in c(\cT_X)) = p_n(a+b)\cdot p_{a+b}(b).
\end{equation}
Thus, substituting (\ref{ppeq3}) and  (\ref{ppeq4}) and the equality $\PP(\cF_2) = p_n(b)$  into  (\ref{ppeq2}), we obtain:
$$p_n(A,B) =p_{n-b}(a)\cdot p_n(b) -p_n(a+b)\cdot p_{a+b}(b) + \hat{p}_n(a,b).$$
Case 5 now follows from Lemmas  \ref{lemclus}, \ref{helpslem}.
Case 6 follows from the compatibility condition (\ref{nesteq}) for clades.
 \hfill$\Box$

\bigskip
We now ask whether the events `$A$ is a clade' and `$B$ is a clade' are positively or negatively correlated under the YHK process.  Let $X_A$ (respectively $X_B$) be the
Bernoulli (0,1) random variables that take the value 1 if $A$ (respectively $B$) is a clade of a YHK tree $\cT$ on $X_n$ and let $\rho_n(A,B)$ denote the
correlation coefficient of these two random variables, which is given by:
$$\rho_n(A,B) = \frac{p_n(A,B)-p_n(A)p_n(B)}{\sqrt{p_n(A)(1-p_n(A))p_n(B)(1-p_n(B))}}.$$

\begin{corollary}
For any two strict subsets $A,B$ of $X$, the correlation $\rho_n(A,B)$ is:
\begin{itemize}
\item strictly negative, if $A, B$ are not compatible, and undefined if $|A|=1$ or $|B|=1$.
\item strictly positive, otherwise.
\end{itemize}
\end{corollary}
{\em Proof:}
If $A$ and $B$ are not compatible, then $p_n(A,B)=0$ but both $p_n(A)$ and $p_n(B)$ are greater than zero, and so $\rho_n(A,B)<0$.
If $|A|=1$ then $p_n(A)=1$ and $p_n(A,B)= p_n(B)$ (regardless of whether $A$ is a subset of $B$ or is disjoint from $B$). Thus
the numerator and denominator of $p_n(A,B)$ are both zero. A similar argument holds if $|B|=1$.

In the remaining cases, we consider the ratio $p_n(A,B)/(p_n(A)p_n(B))$.
For example, in Case 2, we have:
$$\frac{p_n(A,B)}{p_n(A)\cdot p_n(B)} = \frac{(n-1)\cdots(n-a+1)}{(b-1) \cdots (b-a+1)}.$$
This is strictly $>1$ since $\frac{n-1}{b-1}>1, \cdots, \frac{n-a+1}{b-a+1}>1.$
Similar arguments apply in the other cases; however Case 5 requires some detailed algebraic manipulation.
\hfill$\Box$

\begin{figure}[ht]
\begin{center}
\resizebox{12cm}{!}{
\includegraphics{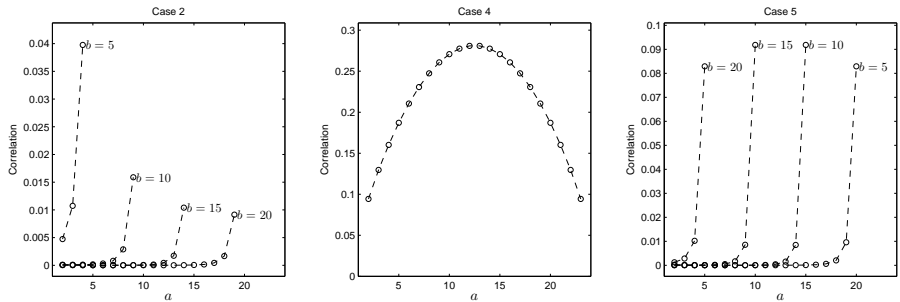}
}
\end{center}
\caption{Graphs of $\rho_{n}(A,B)$ for $n=25$, in Cases 2,4 and 5, with $a=|A|$ and $b=|B|$.}
\label{figrho}
\end{figure}

Fig.~\ref{figrho} illustrates the correlation coefficient $\rho_n(A,B)$ for $n=25$ in the Cases 2, 4 and 5. Notice that, the correlation is typically much smaller in Cases 2 and 5 than for Case 4.

\section{Extension to partitions of $X$.}

 Suppose that the collection of sets $A_1, A_2, \ldots, A_k$ forms a partition of $X$, and let $a_i = |A_i|$, for $i=1,\ldots, k$, so that
 $n=|X| = \sum_{i=1}^k a_i$.
   For a rooted YHK tree $\cT$,
 let  $p(a_1, \ldots, a_k)$ be the probability that $A_1, A_2, \ldots, A_k$ are clades of $\cT$ (this probability depends only
 on the cardinality of the sets by the exchangeability property).  For example, p(2,2,2)=2/225, and
 from Lemma \ref{baslem}, we have:
 $p(a_1, a_2) = \frac{2}{a_1+a_2-1}\binom{a_1+a_2}{a_1}^{-1}.$
Our aim in this section is to generalize this to larger values of $k$.  In order to do so, we describe a new result for the Yule model, which requires a further
definition.

For a rooted YHK tree $\cT$, and a rooted phylogenetic tree
$T_k$  with leaf set $\{1,\ldots, k\}$,
 let  $p(a_1, \ldots, a_k; T_k)$ be the probability that $A_1, A_2, \ldots, A_k$ are clades of $\cT$ and that $T_k$ is the tree obtained from $\cT$ by
 replacing each clade $A_i$ by a single leaf labelled $i$, for $i=1, \ldots, k$.   Let $\cI(T_k)$ denote the set of interior vertices of $T_k$.

 \begin{theorem}
 \label{kcasethm}
 For $k>1$, we have:
 \begin{itemize}
 \item[(i)] $$p(a_1, \ldots, a_k; T_k) = \frac{2^{k-1}\prod_{i=1}^k a_i!}{n!} \prod_{v \in \cI(T_k)} \left(\frac{1}{\sum_{i=1}^k a_iI_v(A_i) -1} \right),$$ where
$I_v(A_i)$ is the indicator variable that takes the value of $1$ if $i$ lies below $v$ in $T_k$ and $0$ otherwise.
\item[(ii)]  $$p(a_1, \ldots, a_k)  = \sum_{T_k} p(a_1, \ldots, a_k; T_k),$$ where the summation is over all distinct rooted binary phylogenetic trees on leaf set $\{1, \ldots, k\}$.
\end{itemize}
\end{theorem}
{\em Proof:}
We prove the result by induction on $k$.  For $k=2$, Lemma \ref{baslem} gives
$p(a_1, a_2; T_2) =\hat{p}_n(a_1, a_2) = \frac{2}{n-1}\binom{n}{a}^{-1}$, where $n=a_1+a_2$, which agrees with
the expression given in part (i) with $k=2$.

Now suppose that part (i) holds whenever $k$ is less or equal to $m \geq 2$; we will show that it also holds when $k=m+1$.  Thus, suppose we have a collection  $C= \{A_1, \ldots, A_{m+1}\}$ that partitions $X$, and also have a rooted binary
phylogenetic tree $T_{m+1}$ on leaf set $\{1, \ldots, m+1\}$.  Then $T_{m+1}$  has a cherry (two leaves adjacent to the same vertex). Without loss of generality (by re-ordering the sets if necessary), we may suppose that these two leaves are $m$ and $m+1$.
Consider the collection of $m$ sets obtained from $C$ by replacing $A_m$ and $A_{m+1}$ by
their union, and let $T'$ be the tree obtained from $T_{m+1}$ by deleting the leaves $m$ and $m+1$ along with their incident edges and labelling the exposed vertex by $m$. Notice that $T'$  rooted binary phylogenetic tree  that has leaf set $\{1, \ldots, m\}$.
By the exchangeability and group elimination (via sampling consistency) properties we have, for $a'_m:=a_m+a_{m+1}$, the following identity:
$$p(a_1, \ldots, a_{m+1}; T_{m+1}) = p(a_1, \ldots, a'_m; T') \cdot \hat{p}_{a'_m}(a_m, a_{m+1}),$$
where $\hat{p}_{a'_m}(a_m, a_{m+1})$ is the probability that a Yule tree on leaf set $A_m \cup A_{m+1}$ has $A_m$ and $A_{m+1}$ as sister
(and thus maximal) clades.
Applying the induction hypothesis for the first term on the right-hand side of this equation, namely $p(a_1, \ldots, a'_m; T')$, and
applying Lemma~\ref{baslem} for the second term, and collecting terms, leads to the expression in Part (i) for $k=m+1$ and thereby justifies the induction step.

Part (ii) follows by observing that each tree $\cT$ that has $A_1, \ldots, A_k$ as clades has one (and only one) associated tree $T_k$, and so these trees provide a partition of the
event for which the probability is given by $p(a_1, \ldots, a_k)$.
\hfill$\Box$

As an illustration of Theorem~\ref{kcasethm}, we have the following result for $k=3$:
$$p(a_1, a_2, a_3) = \frac{4�a_1!a_2!a_3!}{n!(n-1)}\left[\sum_{i=1}^3 \frac{1}{n-a_i-1}\right],$$
where $n=a_1+a_2+a_3$.

We note that, as well as  being  a generalization of Lemma \ref{baslem} to $k>2$, Theorem~\ref{kcasethm}(i) also generalizes the classic result that the probability that a YHK tree
$\cT$ has a given tree topology $T_k$ is $\frac{2^{n-1}}{k!}\prod_{v \in \cI(T_k)}\left(\frac{1}{n_v-1}\right)$, where $n_v$ is the number of leaves of $T_k$ below $v$ (see \cite{brown} or \cite{sem}). This can be seen by setting  $a_1=a_2=\cdots =a_n=1$ in Theorem ~\ref{kcasethm}(i) .

\section{Extension to unrooted trees}

If we suppress the root $\rho$ of a rooted binary phylogenetic $X$--tree $T$, we obtain an unrooted binary phylogenetic $X$--tree, which we will denote as $T^{-\rho}$ (as shown in Fig.~\ref{fig1}(b)). Following \cite{wilk}, (see also \cite{lap}) we say that a subset $A$ of $X$ is a {\em clan} of an unrooted phylogenetic $X$--tree $T'$ if $A|X-A$ is a split of $T'$. Note that any clade of the rooted tree $T$ becomes a clan of $T^{-\rho}$. However, this latter tree also has additional clans that do not correspond to a clade of $T$. The precise relationship is given as follows:

\begin{lemma}
\label{linklem}
Given a rooted binary $X$--tree, $T$, a set $A$ is a clan of $T^{-\rho}$ if and only if either $A$ is a clade of $T$ or $X-A$ is a clade of $T$.
\end{lemma}

\begin{figure}[ht]
\resizebox{12cm}{!}{
\includegraphics{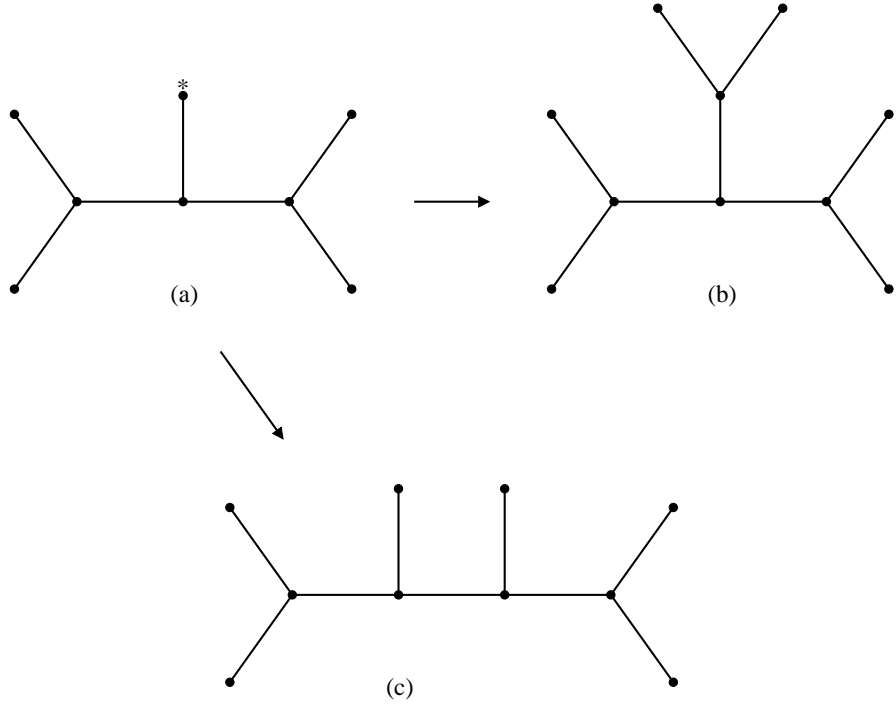}
}
\caption{Only one unrooted binary tree shape is possible with five leaves (a), but two are possible with six leaves (b, c).  If the `central' leaf (*) of tree $a$ is split to form two leaves, then we obtain tree shape (b), while splitting any one of the remaining four leaves produces tree shape (c). Thus, tree shape (b) has probability $1/5$. Since there are $6!/3!2^3= 15$ distinct ways to label its leaves, each of the resulting phylogenetic trees has probability $1/75$. By contrast, any phylogenetic tree of shape (c) has probability $4/5 \times 1/90 = 2/225$.}
\label{fig3}
\end{figure}

Now suppose the  rooted phylogenetic tree $T$ is generated under the YHK process. Then we obtain an induced probability for the
unrooted tree $T^{-\rho}$. Note that the same unrooted tree can arise from different rootings.  This probability distribution on unrooted phylogenetic trees can also be described directly as a Yule-type process on unrooted trees
in which, at each stage, a leaf is selected uniformly at random and a new leaf (with a random label) is attached to its incident edge (see e.g. \cite{penny}).  Fig.~\ref{fig3} illustrates how different leaf choices  in this process lead to different  shapes of unrooted trees.

For a strict non-empty subset $A$ of $X_n$, let $q_n(A)$ be the probability that $A$ is a clan of the unrooted YHK tree on leaf set $X_n$; by (EP) this depends only on
$a =|A|$ and $n$  so we will also write it as $q_n(a)$.

\begin{lemma}
\label{qlem}
\mbox{ }
$$q_n(a) = 2n\left[\frac{1}{a(a+1)} + \frac{1}{b(b+1)}- \frac{1}{(n-1)n}\right] \binom{n}{a}^{-1},$$
where $a=|A|, b = n-a$.
\end{lemma}
{\em Proof:}
By Lemma~\ref{linklem}, we have:
$$q_n(A)= p_n(A)+p_n(X-A) - p_n(A, X-A).$$
Applying Lemmas~\ref{lemclus} and \ref{baslem}, noting that $p_n(A, X-A) = \hat{p}_n(A, X-A)$, leads to the claimed equation.
\hfill$\Box$
\bigskip

Now consider two disjoint subsets $A$ and $B$ of $X$, and let $q_n(A,B)$ be the probability that $A$ and $B$ are both clans of the unrooted YHK tree on leaf set $X_n$.
By (EP), this probability depends only on $a = |A|, b=|B|$ and $n$, and so we will denote it as $q_n(a,b)$.
As an example, we have:
$$q_6(2,2) = 7/225.$$
To see this, observe that if we take (say) $A= \{1,2\}, B=\{3,4\}$ then, referring to Fig.~\ref{fig3}, there is just one tree of shape (b) and two of shape (c) that has both clans $A$ and $B$.  Thus, $q_6(2,2) = 1\times \frac{1}{75} + 2 \times \frac{2}{225}.$
We now give an exact analytical formula for $q_n(a,b)$.

\begin{theorem}
\label{unroothm}
\mbox{}
\begin{itemize}
\item[(i)]
If $a+b=n$, then:
$$q_n(a,b) = q_{a+b}(A) = \frac{2a!b!}{(a+b-1)!}\left[\frac{1}{a(a+1)}+ \frac{1}{b(b+1)}-\frac{1}{(a+b)(a+b-1)}\right].$$
\item[(ii)]
If $a+b<n$ then:
$$q_n(a,b) =  r_n(a,b) +  R_n(a, n-b) + R_n(b, n-a)  - \hat{p}_n(b, n-b)p_{n-b}(a) -\hat{p}_n(a, n-a)p_{n-a}(b),$$
where the first three quantities are as given in Theorem~\ref{main} (Cases 2, 3 and 5), while the last two terms are given by
Lemmas \ref{lemclus} and \ref{baslem}.

\end{itemize}
\end{theorem}
{\em Proof:}
Part (i) follows from Lemma~\ref{qlem}, noting that $n=a+b$.  For part (ii),  Lemma~\ref{linklem} implies that $A$ and $B$ are clans of $T^{-\rho}$ precisely if one of the following three events occur:
\begin{itemize}
\item[(a)] $A$ and $B$ are clades of $T$;
\item[(b)]  $A$ and $X-B$ are clades of $T$, but $B$ is not a clade of $T$;
\item[(c)] $B$ and $X-A$ are clades of $T$, but $A$ is not a clade of $T$;
\end{itemize}
(Note that $X-A$ and $X-B$ cannot both be clades of $T$, by the compatibility condition (\ref{nesteq})
 since $(X-A) \cap (X-B) \neq \emptyset$ by the assumption that $a+b<n$, and since $X-A$ neither contains nor is contained in $X-B$).
 Moreover, the three events (a), (b), (c) are mutually exclusive, by virtue of the
the assumption that $A, B$ are disjoint and their union is a strict subset of $X$.
The probability of Event (a) is $r_n(a,b)$, while the probability of Event (b) is $R_n(a,n-b) - \hat{p}_n(b, n-b)p_{n-b}(a)$ since the first term
is the probability that $A$ and $X-B$ are clades of $\cT$, and $\hat{p}_n(b, n-b)p_{n-b}(a)$ is the probability that $A, X-B$ and $B$ are clades of $\cT$.
Similarly, $R_n(b, n-a) - \hat{p}_n(a, n-a)p_{n-a}(b)$ is the probability of Event (c). The result now follows by adding the probabilities of these three
mutually exclusive events.
\hfill$\Box$

\subsection{Extensions of the clan condition (I)}

For a pair $A,B$ of disjoint subsets of $X$ a weaker condition than requiring that $A$ and $B$ are both clans of $\cT^{-\rho}$ is simply to require that at least one edge of this tree separates $A$ from $B$. Let $Q_n(A,B)$ be the probability of this event for an unrooted YHK tree on the leaf set $X_n$.  Then we have the following result, which follows from the sampling consistency (SC) property applied in the unrooted setting.
\begin{equation}
\label{Qqeq}
Q_n(A,B) =  q_{a+b}(A),
\end{equation}
where $q_{a+b}(A)$ is given by Theorem~\ref{unroothm}(i).

\subsection{Extensions of the clan condition (II)}
We now describe a second extension.   Suppose $A_1, A_2, \ldots, A_k$ partition $X$, and, as usual, let $a_i = |A_i|$.  For an unrooted YHK tree $\cT$ let
$q(a_1, \ldots, a_k)$ be the probability that $A_1, A_2, \ldots, A_k$ are clans of $\cT$ and let $q'(a_1, \ldots, a_k)$ be the probability that $A_1, A_2, \ldots, A_k$ are convex on $\cT$ (that is, the minimal subtree connecting
the leaves in $A_i$ is vertex disjoint from the minimal subtree connecting the leaves in $A_j$ for all pairs $i,j$; see \cite{sem} for further details and the biological significance of convexity).

We have calculated $q$ when $k=2$ above (and  $q' = q$ in this case). We turn now to the next case of of interest, $k=3$, where, for example, we  have:
$$q(2,2,2)=1/75, \mbox{ and }  q'(2,2,2)=1/15.$$  The following result provides an exact formulae for these two quantities for arbitrary $(a_1, a_2, a_3)$.

\begin{theorem}
Let $n=a_1+a_2+a_3$. Then:
\begin{itemize}

\item[(ii)]  $q(a_1, a_2, a_3)=\frac{4a_1!a_2!a_3!}{(n-1)!}\left[\sum_{i=1}^3\frac{1}{(n-a_i)((n-a_i)^2-1)} \right].$
\item[(ii)]  $q'(a_1, a_2, a_3)=q_n(a_1,a_2)+q_n(a_1,a_3)+q_n(a_2,a_3)-2q(a_1,a_2,a_3),$
where $q_n(a_i, a_j)$ is given in Theorem~\ref{unroothm}(ii), and $q(a_1, a_2, a_3)$ is from part (ii).
\end{itemize}
\end{theorem}
{\em Proof:}  For part (i), the event that  $A_1, A_2$ and $A_3$ (which partition $X$) are clans of $\cT^{-\rho}$ is the union of three disjoint events $E_{jk}$ over the three choices of $\{j,k\} \in \{\{1,2\}, \{1,3\}, \{2,3\}\}$, where
$E_{jk}$ is the event that the union of two of the sets -- say $A_j$ and $A_k$ --  must be a clade of $\cT$, and that this clade has  maximal clades $A_j$ and $A_k$. The exchangeability and group elimination conditions then give:
$$q(a_1, a_2, a_3) = \PP(E_{12}) + \PP(E_{13}) + \PP(E_{23}) =  \sum_{i=1}^3 p_n(n-a_i)\cdot \hat{p}_{a_j+a_k}(a_j, a_k),$$
where $\{a_i,a_j, a_k\} = \{1,2,3\}$ in the term on the right-hand side of this last equation.
By Lemmas \ref{lemclus} and \ref{baslem},  this gives:
$$q(a_1, a_2, a_3) = \sum_{i=1}^3 \frac{2n}{(n-a_i)(n-a_i+1)} \frac{(n-a_i)!a_i!}{n!}\cdot \frac{2}{(n-a_i-1)}\frac{a_j!a_k!}{(n-a_i)!}$$
which simplifies to the expression given in (ii).

For part (ii), the event that $A_1, A_2$ and $A_3$ are convex on $\cT^{-\rho}$ is the union of three (non-disjoint!) events $E'_{jk}$ over the three choices of $\{j,k\} \in \{\{1,2\}, \{1,3\}, \{2,3\}\}$, where
$E'_{jk}$ is the event that two of the sets -- say $A_j$ and $A_k$ -- are clans of $\cT^{-\rho}$. Note that the intersection of any two (or three) of these three events is simply the event that all three sets are
clans of $\cT$, which was dealt with in part (i). Thus, by the principle of inclusion and exclusion, we have:
$$q'(a_1, a_2, a_3) = \PP(E'_{12})+\PP(E'_{13})+\PP(E'_{23}) - 2q(a_1, a_2, a_3)$$ and the
result in part (iii) now follows.

\hfill$\Box$

Deriving explicit formulae for $q(a_1, \ldots, a_k)$ and $q'(a_1, \ldots, a_k)$ for $k>3$ is, in principle, possible but the formulae quickly become increasingly complex.

\subsection{Extensions of the clan condition (III)}
A third extension is to consider the probability $Q_n(A_1, A_2)$  that two sets $A_1, A_2$ are clans of a YHK tree on $n$ leaves when these two sets are {\bf not} disjoint.  For this setting we have the following result.

 \begin{proposition}
 \label{connectprop}
 Suppose $A_1, A_2$ are non-disjoint subsets of $X$, and $a_i = |A_i|$.
 \begin{itemize}
 \item[(i)] If $A_1 \subset A_2$, then:
 $$Q_n(A_1, A_2) = q_n(a_1, n-a_2),$$ where
  $q_n(*,*)$ is given by Theorem  \ref{unroothm}.   Similarly, if $A_2 \subset A_1$ then
  $Q_n(A_1, A_2) = q_n(n-a_1, a_2).$
 \item[(ii)]  Otherwise, if neither set $A_1, A_2$ is a subset of the other, then:
 $$Q_n(A_1, A_2) = \begin{cases}
 q_n(a_1-a_{12}, a_2-a_{12}), &\text{ if } A_1 \cup A_2 = X;\\
 0, &\text{ otherwise. }
\end{cases}
 $$
 where $a_{12}= |A_1\cap A_2|$, and $q_n(*,*)$ is given by Theorem \ref{unroothm}.
 \end{itemize}
 \end{proposition}
 {\em Proof:}
First observe that if $A_1 \subset A_2$ then $A_1$ and $A_2$ are clans of an unrooted phylogenetic $X$--tree $T$ if and only if $A_1$ and $X-A_2$ are clans of $T$. Noting that these are disjoint sets, the first part of Proposition \ref{connectprop} follows from
Theorem  \ref{unroothm}.
For the second case, where neither set $A_1, A_2$ is a subset of the other, first observe that in order for $A_1$ and $A_2$ to be clans of the same unrooted phylogenetic $X$--tree $T$ a necessary condition is that
$A_1 \cup A_2 = X$. Moreover, under this condition, $A_1$ and $A_2$ are clans of $T$ if and only if
$A_1-A_1\cap A_2$ and $A_2- A_1\cap A_2$ are clans of $T$;  as these are disjoint sets, the second part of Proposition~\ref{connectprop} follows from Theorem \ref{unroothm}.
 \hfill$\Box$

\section{Discussion}

The arguments we have used in our analysis have primarily relied on repeated application of the properties of exchangeability (EP) and group elimination (GE) (or its corollary, sampling consistency (SC)) for the YHK model, together with Lemmas \ref{lemclus} and \ref{baslem}.   However other natural models for trees can also satisfy some of these properties. Indeed the distribution that assigns each rooted binary phylogenetic tree on $X_n$ the
same probability (sometimes known as the `Proportional to Distinguishable Arrangements', or PDA model)  satisfies both (EP) and (GE) \cite{ald}.  This suggests that by finding and applying the corresponding
results to Lemma \ref{lemclus} and \ref{baslem} for the PDA model, one could develop a parallel line of results for the PDA model to most of the analysis we have provided in this paper for the YHK model.

 Unfortunately
only one other model, apart from PDA and YHK,  is known to satisfy both (EP) and (GE) and this model is not of biological interest, as it only generates pectinate (comb-like) tree shapes.  Aldous \cite{ald} has conjectured
that these are the {\em only} three distributions on rooted binary phylogenetic trees that that satisfy both (EP) and (GE).  Nonetheless, it may be of interest to explore models that satisfy weakened assumptions - for example, (EP) and (SC), or just (EP).

Even with (EP) alone, one can devise meaningful statistical significance tests.  For example, suppose $N$ taxa include one or more particular (disjoint) subsets (different `types' of taxa) $A_1, A_2, \ldots, A_k$, where $k \geq 1$.
Consider any model for generating a rooted binary tree that satisfies the exchangeability property (EP), and let  $p_n$ be the probability that a tree on this set of taxa as leaves, generated under this model,  has at least one clade of size at least $n$ consisting of just one type (i.e. all leaves in the clade are a subset of one of the sets $A_1, \ldots, A_k$). Then we have the following result, the proof of which is given in the Appendix.

\begin{proposition}
\label{proptest}
For any probability distribution on rooted binary trees satisfying the exchangeability property (EP), we have:
$$p_n  \leq \sum_{i=1}^k \sum_{m=n}^{a_i}   \frac{\binom{a_i}{m}}{\binom{N-1}{m-1}},$$
where $a_i = |A_i|$.
\end{proposition}
As a simple example, suppose we have $N=40$ taxa, including two disjoint groups, each containing  six taxa. For a tree generated under any model that satisfies the exchangeability property, the probability that this tree would contain
a clade of size four of larger consisting entirely of taxa from one of the two groups is, at most:
$$2 \cdot \left( \frac{\binom{6}{4}}{\binom{39}{3}} +  \frac{\binom{6}{5}}{\binom{39}{4}} +  \frac{\binom{6}{6}}{\binom{39}{5}} \right) < 0.005.$$

\section{Acknowledgements}

We thank the Royal Society of NZ (Marsden Fund and James Cook Fellowship) and the Allan Wilson Centre for Molecular Ecology and Evolution for funding this work.
We also thank Arne Mooers for comments that motivated the development of  Proposition \ref{proptest}.

\section{Appendix: Proof of Proposition~\ref{proptest}}

Let $X_{m,i}$ be the number of clades of size $m$ in the randomly-generated tree that has the property that the taxa are all of type $A_i$, and let
$X:= \sum_{i=1}^k \sum_{m=n}^{a_i} X_{m,i}.$
Then $p_n=\PP(X>0)$. Since $X$ is a non-negative integer random variable, we have:
\begin{equation}
\label{pp}
\PP(X>0) \leq \EE[X].
\end{equation}
By linearity of expectation we have:
\begin{equation}
\label{line}
\EE[X] =  \sum_{i=1}^k \sum_{m=n}^{a_i} \EE[X_{m,i}].
\end{equation}
Moreover:
\begin{equation}
\label{ee}
\EE[X_{m,i}]= \sum_{t} \EE[X_{m,i}|t]\PP(t),
\end{equation}
where the summation is over all binary tree shapes on the given leaf set of size $N$, $\EE[X_{m,i}|t]$ is the conditional expectation of $X_{m,i}$ given that $t$ is the tree shape
generated by the random speciation process, and $\PP(t)$ is the probability of generating tree shape $t$.
For any given the tree shape $t$:
\begin{equation}
\label{pp2}
\EE[X_{m,i}|t] = \sum_{v: n_v=m} \EE[I_{v,i}|t],
\end{equation}
where the summation is over all the interior vertices of $t$ for which the number of leaves below $v$ ($n_v$) is $m$, and where
$I_{v,i}$ is the binary random variable that takes the value $1$ precisely if all the leaves below $v$ are of type $A_i$, and $I_{v,i}=0$ otherwise.
Now, by exchangeability, we have the following identity for any vertex $v$ of $t$ with $n_v = m$:
\begin{equation}
\label{pp3}
\EE[I_{v,i}|t] =  \PP(I_{v,i}=1|t)= \frac{\binom{a_i}{m}}{\binom{N}{m}},
\end{equation}
Now any tree shape on $N$ leaves has, at most, $N/m$ vertices $v$ for which $n_v =m$, and so we obtain, from (\ref{pp2}) and (\ref{pp3}),
$\EE[X_{m,i}|t]  \leq  \frac{N}{m} \cdot \frac{\binom{a_i}{m}}{\binom{N}{m}} =  \frac{\binom{a_i}{m}}{\binom{N-1}{m-1}}.$
Since this inequality holds for all tree shapes $t$, Equation (\ref{ee}) implies that:
$\EE[X_{m,i}] \leq  \frac{\binom{a_i}{m}}{\binom{N-1}{m-1}}.$
The expression for $p_n$ now follows from Equations (\ref{pp}) and  (\ref{line}).

\hfill$\Box$

\end{document}